# Comment: The Importance of Jeffreys's Legacy

**Robert Kass**

*Abstract.* *Theory of Probability* is distinguished by several high-level philosophical attitudes, some stressed by Jeffreys, some implicit. By reviewing these we may recognize the importance in this work in the historical development of statistics.

Jeffreys is one of the major figures in the history of statistics, and *Theory of Probability* is his chief work on the subject. It is wonderful that Robert, Chopin, and Rousseau (RCR) have devoted so much effort to pouring through the book. Their insights will be appreciated by all future readers.

The two elements of Bayesian analysis most strongly attributed to Jeffreys are Bayes factors and the selection of priors by formal rules. My own understanding of these subjects was embedded in a pair of reviews roughly 15 years ago (Kass and Raftery, 1995; Kass and Wasserman, 1996). Now, however, my priorities have evolved—I have spent much of the past 10 years worrying about the application of statistics to problems in neuroscience, and trying to identify the most important lessons from our discipline that should be passed on to budding data analysts. I would like to offer a few comments on Jeffreys's legacy from this current perspective. Or, perhaps my aim is better communicated by asking, How should our legacy be informed by Jeffreys's legacy? Before getting to this high-level question I would like to make one technical remark.

*Robert Kass is Professor, Department of Statistics, Machine Learning Department, and Center for the Neural Basis of Cognition, Carnegie Mellon University, Pittsburgh, Pennsylvania 15213, USA e-mail: kass@stat.cmu.edu.*

## 1. GEOMETRY

It is worth noting Jeffreys's clear geometrical thinking in his choice of general prior, discussed at the beginning of RCR's Section 4.7. The lurking differential geometry is subtle, as he says only that Hellinger distance and Kullback–Leibler divergence "have the form of the square of an element of distance in curvilinear coordinates," but, in addition, his notation $g_{ik}$ for the $(i,k)$ element of the Fisher information matrix is the standard notation of his day for the elements of the matrix representation for a Riemannian metric. It was obvious that the Riemannian natural volume element—the determinant of the matrix representation of the metric—would be invariant and, as argued in Kass (1989), this provides a bit of intuition. In addition, basic to Jeffreys's treatment of Bayes factors was his use of orthogonal parameters (see Section 5.01 of *Theory of Probability* and Section 6.1 of RCR)—which are "orthogonal" in the sense of differential geometry. Furthermore, his use of his general prior for Bayes factors, discussed in RCR Section 6.4, is again intelligible as a prior (approximately) on the resulting Riemannian distance (the "information distance" discussed in Kass, 1989). I have always assumed this was an important part of Jeffreys's thought process.

## 2. THE BAYESIAN ENGINE

As I look back again, now, on *Theory of Probability* I find four particularly striking features.

First, it treated a wide variety of problems, many of which continue to be of interest. [A list of a dozen such problems appeared in Kass (1991) as part of a special issue of *Chance* devoted to the 100th anniversary of Jeffreys's birth.] In this integration of theory and practice it became a model text. Indeed, in the intervening years there has been an unfortunate bifurcation of theory and practice, so that theoretical texts rarely give the kind of attention to practical problems that Jeffreys did.

Second, it relied on first-order approximation, via Laplace's method, especially to center the posterior at the MLE (though modern applications often use the posterior mode). This is noteworthy, in part, because it played a fundamental role in his views on selecting priors. Jeffreys admitted the choice of prior was somewhat arbitrary, but he pointed out that asymptotic considerations made this degree of arbitrariness a rather minor practical difficulty. Over the years there has been some misunderstanding of Jeffreys's point of view because it changed over time in response to critics—this is one reason it is worth examining the multiple editions of his book. (See Kass and Wasserman, 1995, Section 2.) Furthermore, we see in Jeffreys's use of Laplace's method the germs of Bayesian computation: he recognized, more clearly than many subsequent researchers, who were concerned with exact results, how Bayes's theorem could be applied in a wide range of analytically intractable problems. And, of course, first-order asymptotics brought Jeffreys's methods into close agreement with Fisher's. In the preface of the first edition of *Theory of Probability* Jeffreys stated, "There is, on the whole, a very good agreement with the recommendations made in statistical practice." In my view it is worth emphasizing the use of first-order asymptotics because much elaborate, painstaking statistical work ends up being useful in scientific inference mainly in its ability to provide a well-founded estimate and standard error. In contemplating the practical value of his treatise, Jeffreys recognized this as well when he said, in the preface to its third edition, "There is a decided improvement in the willingness of physicists to estimate uncertainties of their results properly, and I suppose that I can claim some of the credit for this."

A third striking high-level feature of *Theory of Probability* is its championing of posterior probabilities of hypotheses (Bayes factors), which made a huge contribution to epistemology. Emanating from his early work with Dorothy Wrinch, this was Jeffreys's main motivation for writing the book. In the preface to the first edition he wrote,

> In opposition to the statistical school, [physicists] and some other scientists are liable to say that a hypothesis is definitely proved by observation, which is certainly a logical fallacy; most statisticians appear to regard observations as a basis for possibly rejecting hypotheses, but in no case for supporting them. The latter attitude, if adopted consistently, would reduce all inductive inference to guesswork; the former, if adopted consistently, would make it imposssible ever to alter the hypotheses, however badly they agreed with new evidence.... In the present book I ... maintain that the ordinary common-sense notion of probability is capable of precise and consistent treatment when once an adequate language is provided for it. It leads to the results that a precisely stated hypothesis may attain either a high or a negligible probability as a result of observational data.

In showing the world the importance of Bayes' theorem, Jeffreys succeeded spectacularly well. The notion that Bayes' theorem can describe, with beautiful brevity, the way we incorporate information to gain knowledge is very widely accepted—even by those, within and outside of statistics, who are not very fond of Bayesian statistical methods in practice. Laplace made an important start, but Jeffreys took the argument much further by showing how Bayes' theorem may be connected with the fundamental aspirations of science.

Jeffreys's observations opened the door to a unification of epistemology with scientific inference via statistical methodology. This was his great goal, and it has remained a goal of Bayesian "true believers" ever since, even for those who have discarded parts of Jeffreys's philosophy and replaced it with subjectivist foundations. There is an undeniable allure of the power and simplicity of the Bayesian approach—I see it in neuroscience as well as statistics—but, in my opinion, despite all its spendor, the Bayesian approach has not realized the goal of unifying statistical inference, nor is it likely to do so in the forseeable future.

There are many reasons for the failure of the Bayesian grand scheme—in the face of all the Bayesian successes—but one important difficulty is the discrepancy between the conceptual, epistemological use of posterior probabilities and their use in practice. In practice, posterior probabilities are used for model selection (e.g, in reversible jump MCMC) and classification, but they are almost never used in the manner Jeffreys emphasized, namely, to provide evidence in favor of scientific hypotheses. Frequentist significance testing (via bootstrap and permutation tests) is pretty easy, even in relatively complicated situations. Bayesian testing, however, is in one respect difficult even setting aside computational issues: although (as reviewed in Kass and Raftery, 1995) Bayes factors are generally not sensitive to priors on suitably-defined nuisance parameters ("null orthogonal" parameters in the sense of Kass, 1989), they remain sensitive—to first order—to the choice of prior on the parameter being tested. This implies that interpretations such as Jeffreys's, reported by RCR, are contaminated by a constant that does not go away asymptotically. Indeed, this is the reason for the large range of values within Jeffreys's interpretive categories. One may see this as a virtue of the Bayesian approach, that its very ambiguity provides a more thorough ("honest") assessment of evidence, but it does impose a burden on those who wish to make scientific inferences. In some applied situations, the evidence may be "decisive" over a wide enough range of priors to be convincing, and it is possible that continuing research will eventually bring Bayes factors into widespread scientific use. I can report, though, that in neuroscience, despite considerable penetration of Bayesian ideas and Bayesian methods, Bayes factors for scientific hypothesis testing are essentially non-existent. And I have yet to find a good application for them, myself.

Nonetheless, despite the apparent over-reach of the Bayesian aspiration set forth by Jeffreys, these first three components of *Theory of Probability* demonstrated constructively the great power of the Bayesian engine. It had a tremendous influence on the next generation of books, which in turn educated those who became soldiers in the "Bayesian revolution" during the 1990s.

## 3. DECISION THEORY

*Theory of Probability* articulated only one of the two crucial elements in the emergence of modern Bayesian analysis, in statistics and throughout science: Bayes' Theorem as an engine for scientific inference. The second element, the optimality of Bayesian procedures, including especially the optimality of Bayes classifiers, had to wait for Wald (and then others such as Savage and Raiffa and Schlaiffer). It is impossible to sing the praises of *Theory of Probability* without emphasizing the continuing importance of optimality. As RCR point out, Jeffreys did mention the performance of methods, and in fact noted the optimality of the Bayes factor in balancing type I and type II errors, but this appears largely as an afterthought in response to Neyman and Pearson, rather than as the fundamental motivation that frequentist optimality subsequently became.

## 4. STATISTICAL MODELS AND SCIENTIFIC LAWS

The fourth feature of *Theory of Probability* that remains, at least to me, especially important is its identification of scientific laws with statistical models. Jeffreys put it this way:

> A physical law is not an exact prediction, but a statement of the relative probabilities of variations of different amounts.

This passage appeared in the first edition of the book, and was italicized in the second and third editions. The point of view is echoed throughout *Theory of Probability* and it stands in contrast to anything declared by Fisher.

I see this as crucially important to our contemporary situation. In a recent article, Emery Brown and I (Brown and Kass, 2009) noted our disgruntlement with much data analysis we have seen in neuroscience. We put it this way:

> We have seen many highly quantitative researchers trained in physics and engineering, but not statistics, apply sophisticated techniques to analyze their data. These are often appropriate, and sometimes inventive and interesting. In the course of perusing many, many articles over the years, however, we have found ourselves critical of much published work. Starting with vague intuitions, particular algorithms are concocted and applied, from which strong scientific statements are made. Our reaction is too frequently negative: we are dubious of the value of the approach, believing alternatives to be much preferable; or we may concede that a particular method might possibly be a good one, but the authors have done nothing to indicate that it performs well. In specific settings, we often come to the opinion that the science would advance more quickly if the problems were formulated differently—formulated in a manner more familiar to trained statisticians.

This led us to consider what statistical training brings to the table, and we articulated a succinct answer in the form of a pair of dogmas of modern statistical thinking:

1. Statistical models of regularity and variability in data are used to express knowledge and uncertainty about a signal in the presence of noise, via inductive reasoning.
2. Statistical methods may be analyzed to determine how well they are likely to perform.

The claim was not that these two things describe what statisticians do, but rather that they characterize the way they think. The implication, and the main subject of that article, was that we as a profession should conscientiously emphasize these points in our teaching and curriculum development. Here, I would like to add that the first item, stressing statistical models, is central in *Theory of Probability*. Our modern notion of statistical model is much broader than that of Jeffreys, and owes much to Fisher. For example, statistical models are used in bootstrap and permutation tests, as well as a host of nonparametric inference and prediction methods. However, the point that statistical models drive the process remains at the essence of our discipline. (For interesting related remarks see Cox, 2001 and Efron, 2001.) To me, this is the most fundamental message of the *Theory of Probability*.

On the other hand, we can't neglect the second item above, performance of methods. This is equally important to our discipline—yet it is largely absent from *Theory of Probability* (and crucial aspects are also absent from Fisher).

## 5. ON RE-READING JEFFREYS

Because of Jeffreys's emphasis on the connection between scientific laws and statistical models, re-reading *Theory of Probability* always leaves me with a burning question: What is the scientific status of a statistical model? That is, in using a statistical model, to what extent are we making scientific claims?

This foundational issue, at once philosophical and practical, has received considerable discussion over the years, and deserves continued attention. Lehmann and Cox, in special lectures and articles, both pointed out that the extent to which a model is "explanatory" or "empirical" depends on context (Lehmann, 1990; Cox, 1990), and Lehmann cited Kruskal and Neyman (1956) in saying the distinction is not rigid: "[These descriptions] represent somewhat extreme points of a continuum." Freedman repeatedly criticized claims based on statistical models because he felt they were empirical in nature yet were used inappropriately for explanation (e.g., Freedman and Zeisel, 1988, *Statistical Science*). The nature of statistical models is closely related to the nature of scientific models (or theories), which are often regarded as either "real" or "instrumental" (see Stanford, 2006). It is worth asking whether, and how, statistical models are essentially different than other kinds of scientific models.

In discussing the connection between statistical models and scientific laws, some of Jeffreys's favorite examples are chosen for their rhetorical value, such as gravitation. Such nice clean examples where scientific theories are extremely precise are, however, quite rare. Certainly in neuroscience the "theories," even when stated mathematically, are supposed to be provide only rough approximations to reality. The same can be argued in principle in physics, but in the biological realm the "rough approximation" is very rough.

Perhaps all models are similar in their attempt to describe the world, but we in statistics are conscious of their shortcomings, especially when they are statistical models. And perhaps contributions can come from the quintessential statistical attitude, "All models are wrong, but some are useful" (Box, 1979), implemented by stressing essential features captured by models that *do* represent scientific claims, from inessential features that do not.

In any case, as RCR so thoroughly demonstrate, *Theory of Probability* is full of weighty material. Reading it from a contemporary perspective opens up all kinds of questions; questions of detail, and questions about the nature of our discipline. One thing is for sure: it is a landmark in the history of statistics. Reading it helps us better understand the conceptual development of our subject.

## ACKNOWLEDGMENTS

Supported in part by R01 MH064537.

COMMENT 5